\documentstyle[12pt,epsf,aps]{revtex}
\begin{document}
\title{Density-Matrix Approach to a Strongly Coupled\\ Two-Component
Bose-Einstein Condensate}
\author{Andal Narayanan \thanks{e-mail: andal@iiap.ernet.in}\\
Non-Accelerator Particle Physics Group \\
Indian Institute of Astrophysics, Koramangala, Bangalore - 560 034, INDIA\\ and\\
Hema Ramachandran \\
Raman Research Institute, Sadashivnagar, Bangalore - 560 080, INDIA}
\maketitle
\vspace{1 cm}
\begin{abstract}
The time evolution equations 
for average values of population and relative phase
of a strongly coupled two component Bose-Einstein condensate (BEC) 
is derived analytically.
The two components are two hyper-fine states, which are coupled by an external
laser that drives fast Rabi oscillations between these states.
Specifically,
this derivation incorporates 
the two-mode model proposed in $~\cite{jw}$ for the strongly coupled 
hyper-fine states $\left|1,-1\right>$ and $\left|2,1\right>$ of $^{87}Rb$.
The fast Rabi cycles are averaged out and the rate equations so derived 
represent the slow dynamics of the system. 
These include the collapse and revival
of Rabi oscillations and their dependence on detuning and trap
displacement as reported in experiments of  $~\cite{jw}$. 
A procedure for stabilising vortices is also suggested.
\end{abstract}

\section{Introduction}

Observations of Bose-Einstein Condensation in trapped dilute alkali
atoms have opened up both experimental and theoretical challenges
to understand the properties of such systems. The dynamical properties
of a single condensate such as its collective excitations in a trap
due to a time dependent drive has been one such area of research $~\cite{go},~\cite{sa}$.
The experimental realisation of simultaneous creation and confinement of Bose-
Einstein condensates (BECs) in several hyperfine states of a given species of atom
$~\cite{my},~\cite{mi}, ~\cite{mart}$ has led to
investigations on dynamics of two or more overlapping condensates by coupling 
them with an externally applied laser field. 
In particular experimental realisation of binary mixtures of two hyperfine 
states namely $\left|1,-1\right>$ and 
$\left|2,1\right>$ of $^{87}Rb$ has established the 
following properties of this coupled condensed system $~\cite{ma}, ~\cite{ha}$.

\begin{itemize}
\item[{\bf (a)}] These two states have magnetic moments which are same to
the first order. However,due to other small effects such as gravity, the nuclear
magnetic moment and nonlinearity in the Zeeman shifts, the location of minima
for the two states in the trap can be adjusted to be slightly different
or exactly coincident.
\item[{\bf (b)}] Spontaneous inter-conversion from one state to the other is not seen
due to the large difference in internal energies between these two states.
The hyperfine energy is 6.8 GHz. This makes the two condensates 
distinguishable. These can be selectively imaged by choice of an 
appropriate laser.
\item[{\bf (c)}] These condensate states  possess a relative quantum phase which
can be measured. This phase evolves with time the rate being proportional
to the chemical potential difference between the two condensates.
\item[{\bf (d)}] An external laser drive couples these two systems and helps
to coherently transfer population from one state to the other.
\end{itemize}
The mixed condensates thus offer an ideal experimental apparatus to look for 
macroscopic realisations of dynamical effects like
standard Josephson effects $~\cite{zap}$. \\
\indent Theoretical calculations on such Josephson like oscillations in these
coupled Boson Josephson Junctions (BJJ) $~\cite{sme}, ~\cite{rag}$ have shown 
several interesting dynamical 
effects. Recently however $~\cite{jw}$, an experimental observation of an unexpected 
behaviour of these coupled systems was reported. In the limit of sustained and 
large field
strengths of the external coupling laser, that is when $\Omega$, the Rabi frequency,
was five to ten times larger than the trap frequency in
the vertical direction, along which the two
condensates sit displaced $~\cite{hall}$, 
the Rabi oscillations between the hyperfine states
was found to collapse and revive. This occurred on a time scale which is 
large compared to the Rabi period. These slow varying modulations of the fast
Rabi oscillations vanish at zero trap displacement. It was also seen to
vanish when $\delta = 0$, where  $\delta$ = $\omega$ - $\omega_d$ is the detuning
of the external laser frequency ($\omega_d$) from the transition frequency between the
hyper-fine states ($\omega$).
It was shown subsequently in the same paper that this phenomenon was due to a
weak coupling between the low lying motional states of the trap. In particular a
simplified two-mode model was suggested. In this two-mode model the trap ground state
and first excited dipole state were coupled and couplings
to all higher motional states were neglected.
Thus the paper $~\cite{jw}$, demonstrates the possibility of quantum state engineering
of topological excitations, through the interplay between the internal and motional
degrees of freedom of a BEC in a TOP trap. Numerical simulations by solving
the Gross-Pitaeskii (GP) equations for the
coupled system were carried out in $~\cite{jw}$
which reproduced the experimental features. 

In the present paper, we derive the essential experimental features analytically,
using the density matrix approach. Equations for the
fractional population ($Z$) in the hyperfine
states and their relative phase ($\theta$) as a function of time have been obtained. 
Averaging over the Rabi period, these equations represent the slow dynamics
of the system. Collapse and revivals of Rabi oscillations and their
dependence on detuning and trap displacement are seen to match qualitatively
with the experimental results described in $~\cite{jw}$.
A proposal for combining this strongly coupled regime to the weakly coupled
Josephson regime is presented and its role in increasing the stability
of vortices $~\cite{dal}, ~\cite{sin},~\cite{rok}$ is also speculated.

\section{A Density-Matrix method}

The total wave-function of the two-component condensate is denoted by $\psi(r,t)$.
Initially this wave-function just represents the total population ($N_T$) in the
ground state $\left|1,-1\right>$.
An external laser drives the transition from this state to the $\left|2,1\right>$
state coherently. So we can write,
\begin{eqnarray*}
\psi(r,t) = \psi_1(r,t) + \psi_2(r,t) 
\end{eqnarray*}

These two states are of the form $~\cite{jw}$
\begin{eqnarray}
\left|\psi_1\right> = (\alpha_1(t)c_0(t)\left|\phi_0\right>+
\alpha_2(t)d_1(t)\left|\phi_1\right>)\left|1\right>\label{big1}\\
\left|\psi_2\right> = (\alpha_2(t)c_0(t)\left|\phi_0\right>+
\alpha_1^{*}(t)d_1(t)\left|\phi_1\right>)\left|2\right> \label{big2}
\end{eqnarray}
Here $\left|1\right>$ and $\left|2\right>$ refer to the hyperfine (internal)
states and $\left|\phi_0\right>$ and $\left|\phi_1\right>$ refer to the 
motional (external) states and
\begin{center}
\begin{eqnarray*}
\alpha_1(t) = \cos{(\frac{\Omega_{eff}t}{2})}-i(\frac{\delta}{\Omega_{eff}})\sin{(\frac{\Omega_{eff}t}{2})}\\
\alpha_2(t) = -i(\frac{\Omega}{\Omega_{eff}})\sin{(\frac{\Omega_{eff}t}{2})}\\
\Omega_{eff} = \sqrt{\delta^{2} + \Omega^{2}}\\
c_0(t) = \cos{(\frac{\Omega_{01}t}{2})}-i(\frac{\Delta e_{01}}{\Omega_{01}})\sin{(\frac{\Omega_{01}t}{2})}\\
d_1(t) = -i(\frac{2\beta<z>}{\Omega_{01}})\sin{(\frac{\Omega_{01}t}{2})}\\
\Omega_{01} = \sqrt{4\beta^{2}<z>^2+\Delta e_{01}^2}\\
<z>_{ij} = \int{\phi_{i} z \phi_{j} dz}\\
\beta =  \frac{z_{0}\delta\Omega}{\Omega_{eff}^2}\\
\Delta e_{01} = e_{1} - e_{0}
\end{eqnarray*}
\end{center}

Here $\delta$ is the detuning, $\Delta e_{01}$ is the energy difference between
the two trap states, namely the ground state and the first excited dipole 
state. This energy difference is held fixed through the derivation, while
in actuality they will vary with time. $z_{0}$ is the displacement between 
the two condensates and $<z>_{ij}$ is the dipole matrix element which couples
the ground and excited states of the trap. In this derivation $<z>_{ij}$ is held
fixed at $<z>_{01}$. The higher couplings are weak and hence neglected. \\

\indent Taking equations ($~\ref{big1}$) and ($~\ref{big2}$) as starting points, we
point out that the $\left|\psi_{1}\right>$ and $\left|\psi_{2}\right>$ 
individually satisfy the following 
normalized coupled Gross-Pitaeskii ($GP$) equation in the Thomas-Fermi limit 
in an isotropic trap.

\begin{eqnarray}
\frac{d\psi_1(t)}{dt} = \frac{1}{i}\{ [2z_{0}<z>+ &
                \lambda_1 N_1+ N_2 + \frac{\delta}{\omega_z} ]\psi_1(t) + &
                  \frac{\Omega}{\omega_z}\psi_2(t)\} \\
\frac{d\psi_2(t)}{dt} = \frac{1}{i}\{ [-2z_{0}<z>+ &
                \lambda_2 N_2+ N_1 - \frac{\delta}{\omega_z} ]\psi_2(t) + &
                       \frac{\Omega}{\omega_z}\psi_1(t)\}
\end{eqnarray}

\begin{eqnarray*}
N_T  = N_1 + N_2\\
a_{\perp} = \sqrt{\frac{\hbar}{m\omega}}\\
a = a_{11} \sim a_{22} \\ 
\lambda_1 = \frac{a_{11}}{a_{12}} \\
\lambda_2 = \frac{a_{22}}{a_{12}} 
\end{eqnarray*}
In writing the above set of coupled equations, time is in units of trap 
frequency $\omega_z$ and
the spatial variables are scaled by $a_{\perp}$. $a_{11}$ and $a_{22}$ are
respectively the $s$ wave scattering lengths for the two hyperfine species
of the condensates  and $a_{12}$ is the inter-species scattering length.
The various energy terms are given with respect to
the trap energy level $\hbar \omega_z$. 
Due to a near degeneracy of $a_{11}$ and $a_{22}$ scattering lengths
in the case of $^{87}Rb$ the approximation that they are equal can be safely
carried out. The system  is characterized then by a single scattering length $a$ and
a single $\lambda$ = $\lambda_1$ $\sim$ $\lambda_2$.
In deriving these equations, the spatial dependence of the $GP$ wave-functions 
are integrated out (Adiabatic approximation) with respect to the trap wave-functions,
namely $\left|\phi_0(z)\right>$ and $\left|\phi_1(z)\right>$ 
and treated as constants. This 
assumes that the specific changes in the shape of the trap wave-function is 
not playing a major role in the time evolution of the system. 
This assumption is an approximation over what is
actually experimentally seen $~\cite{jw}$ since 
we are only interested in the dynamics
of the fractional population of the hyperfine states. \\
\indent  We proceed to derive the population fractions 
and the relative phase differences between the two hyperfine states by noting that
\begin{eqnarray}
\left|1\right> = \sqrt{N_{1}(t)} e^{i\varphi_1(t)}\\
\left|2\right> = \sqrt{N_{2}(t)} e^{i\varphi_2(t)}\\
Z = \frac{N_1 - N_2}{N_T}\\
\theta = \varphi_2 - \varphi_1\\
\end{eqnarray}
 
By taking appropriate inner products of $\left|\psi_1\right>$
and $\left|\psi_2\right>$ with $\left|\phi_0\right>$
and $\left|\phi_1\right>$  and substituting for
$\left|\psi_1\right>$ and $\left|\psi_2\right>$ in the $GP$ equations,
the form given in equations ($~\ref{big1}$) and ($~\ref{big2}$),
the following rate equations for average values of fractional
population $\left<Z\right>$ 
and relative
phase $\left<\theta\right>$ in the two hyperfine states can be derived.
\begin{equation}
\left<\dot{Z}\right> = \left(\frac{\beta^2 <z>^2}{\Omega_{01}} 
\sin{(\Omega_{01} t)} \right) Z
                     \label{one} \\
\end{equation}
\begin{equation}
\left<\dot{\theta}\right> = 4 \left(\Delta e_{01}  
\frac{\beta^2 <z>^2}{\Omega_{01}^2} \sin^2{(\Omega_{01} t)} \right)
\left[ \frac{1}{\cos^2{(\frac{\Omega_{01} t}{2})} 
+(\frac{\Delta e_{01}}{\Omega_{01}})^2 \sin^2{(\frac{\Omega_{01} t}{2})} } \right]
\label{two}
\end{equation}
In deriving the above equations the
orthonormal relations of trap wave-functions are assumed to be
\begin{eqnarray*}
\left<\phi_i|\phi_j\right> = \delta_{ij} 
\end{eqnarray*}
Equations ($~\ref{one}$) and ($~\ref{two}$) are obtained after averaging over
the fast time period namely that of $\Omega$ in the problem. So these equations
do not explicitly contain $\Omega$.
Analytical expression for $\left<Z\right>$ can then be derived.
\begin{equation}
\left<Z\right> = Z_0 \exp{\left(\frac{\beta^2 <z>^2}{\Omega_{01}^2}
\left[1-\cos{\Omega_{01}t} \right]\right)}
\end{equation}
In the limit of small
detuning $\delta$ ($\delta \ll \Omega$), this is of the form
\begin{equation}
\left<Z\right> = Z_0 \left[1+2 \frac{\beta^2 <z>^2}{\Omega_{01}^2} 
\sin^2{(\frac{\Omega_{01}}{2} t)} \right] \label{main}
\end{equation}
Here $Z_0$ is the initial value of the population at time $t=0$.
\section{Results and Discussions}
The equation ($~\ref{main}$) has all the essential features which are reported
by the experiments and subsequent numerical investigation in $~\cite{jw}$. 
\begin{itemize}
\item[{\bf (1)}] The $\left<Z\right>$, remains a constant when $\delta$ goes to zero
or when $z_{0}$ goes to zero. That is, in the lab frame the fast Rabi oscillations
remain unmodulated.
\item[{\bf (2)}] Equation ($~\ref{main}$) is derived with the implicit assumption
that the condensate has a well defined overall phase which can be measured
relative to a reference. A decoupling of $\left<\theta\right>$
and $\left<Z\right>$ in the time averaged frame over the fast variable
occurs as this phase (both
the slow and fast varying part) averages to zero.
\item[{\bf (3)}] Though the mean-field term does not explicitly enter the expression 
($~\ref{main}$), we can see that the amplitude of modulation increases 
with decreasing $\Delta e_{01}$ - a result which is confirmed by numerical simulations 
in $~\cite{jw}$
which predicts a decrease in $\Delta e_{01}$ for enhanced mean-field effects.
\item[{\bf (4)}] This form of equation ($~\ref{main}$) does not give rise to the 
chaotic behaviour with high values of $z_{0}$ reported in $~\cite{jw}$. 
\end{itemize}

Figure 1 gives a typical curve for the parameters given.\\

\indent In the paper $~\cite{mt}$, a preparation of the vortex mode is presented
in this very two-component system. In this contest it is tempting to
think of the following scheme for stabilising such vortices.
Starting from the strong coupling regime ($\Omega > \omega_z$), the
strength of the external laser field could be gradually decreased over time.
At sometime then, when $\Omega < \omega_z$, 
the population in a particular state 
(which is a combination of motional and internal states)
gets trapped in that state itself, due to Macroscopic Quantum State Trapping
(MQST) effects $~\cite{rag}$, applicable in this regime. More specifically,
if such trapping should occur in the first excited motional state, which
could be a vortex state, then there seems to be a tremendous improvement
achieved in the stability of the vortex. A schematic of such a scenario is
given in Figure 2.
\section{Conclusions}
In this paper we have analytically derived an expression for the
rate of change of fractional 
population of the hyperfine states $\left|1,-1\right>$
and $\left|2,1\right>$ of $^{87}Rb$ 
in the strong coupling regime using the density matrix approach. The derivation
gives analytical results for population evolution after averaging out
the fast dynamical variable namely the Rabi period in the problem.
This derivation is based on a two-mode model for the trap states as proposed
in $~\cite{jw}$.  The main result of our analytical approach is presented
in Equation ($~\ref{main}$).  This equation reproduces most of
the essential features
of the two-mode model presented in $~\cite{jw}$.  Also
the possibility to increase the stability
of vortex state by modulating $\Omega$ is also discussed.

\section{Acknowledgements}
AN wants to acknowledge with pleasure
the useful and interesting time spent at ICTP, Trieste, 
Italy, with Prof. S.R. Shenoy and Dr. A. Smerzi during when some of the techniques
used in these calculations were learnt.

\begin{table}
\caption{Parameter values used to plot Figure 1 are given here}
\begin{tabular}[p]{|lr|lr|} 
$N_T$ & $10^5$ & $\nu_{z}$ & 65 Hz \\
$a_{11}$ & 1.0 $a_{21}$ & $a_{\perp}$ & 1.3 $\mu$m \\
$a_{22}$ & 1.0 $a_{21}$ & $a_{21}$ & 5.5 nm  
\end{tabular}
\end{table}

\begin{figure}[b]
\epsfbox{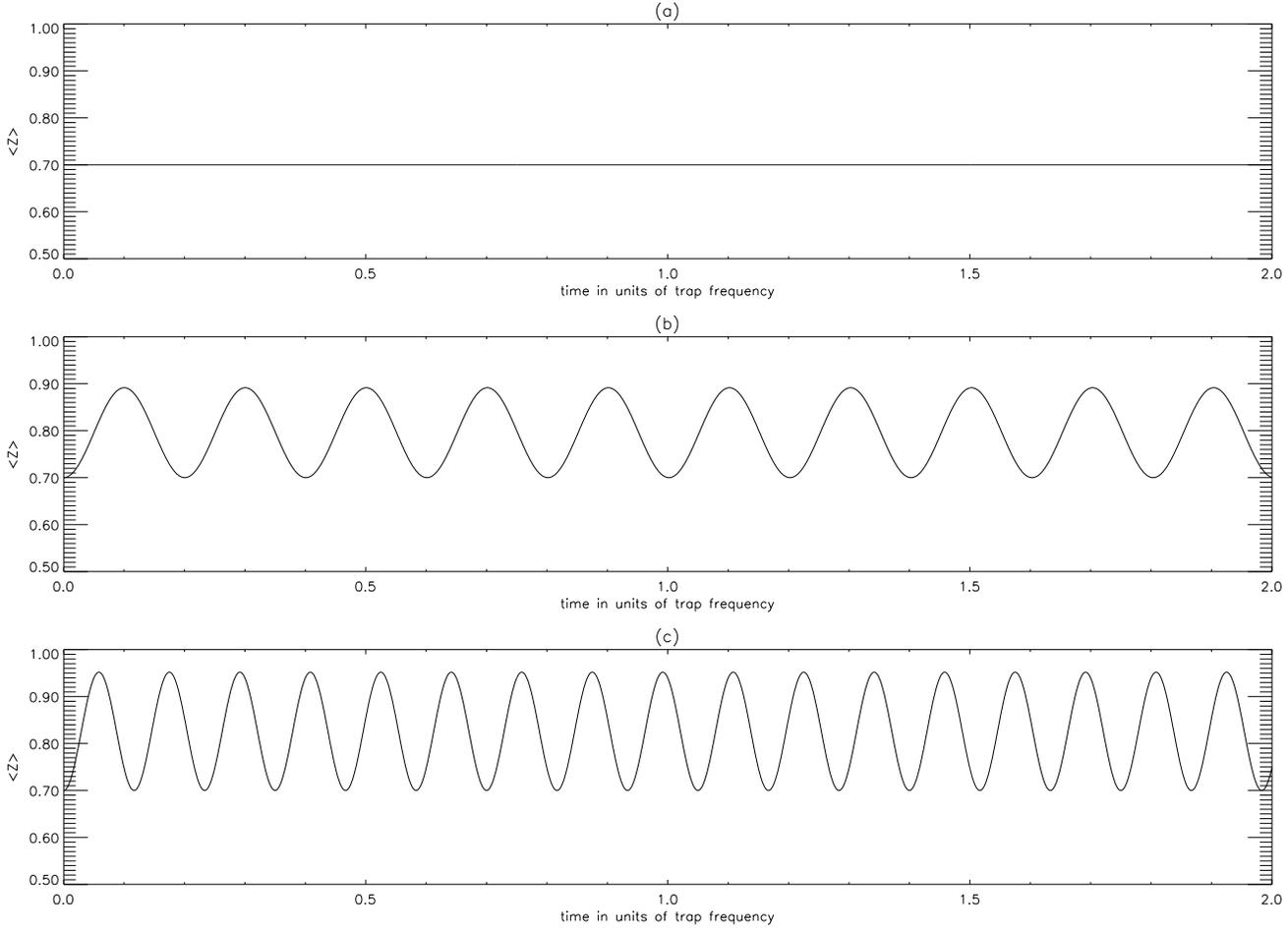}
\caption{This plot shows the change in $\left<Z\right>$ as a function of detuning
$\delta$. The parameters are (a) $\delta = 0$, (b) $\delta = 2\pi*50 Hz$,
(c) $\delta = 2\pi*100 Hz$.
The values of other parameters are given in Table 1.}
\end{figure}

\begin{figure}[b]
\epsfbox{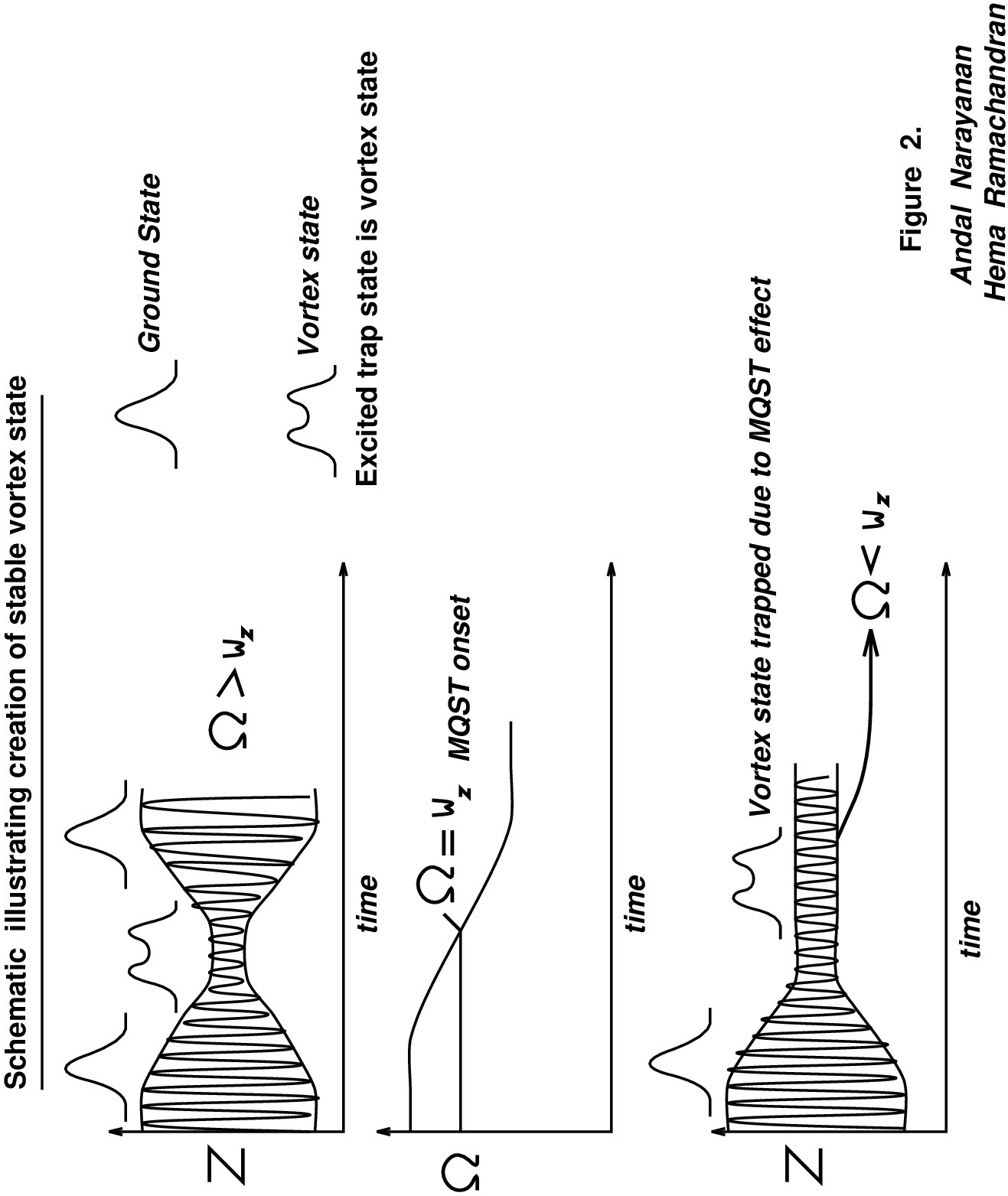}
\end{figure}

\end{document}